\documentclass[aps,prd,twocolumn,superscriptaddress,floatfix]{revtex4-2}

\usepackage{amsmath}
\usepackage{xcolor}

\usepackage{siunitx}
\DeclareSIUnit{\gigayear}{Gyr}
\DeclareSIUnit{\Myr}{Myr}
\usepackage{aas_macros}

\usepackage{amssymb}
\usepackage{graphicx}
\usepackage{dcolumn}
\usepackage{orcidlink}
\usepackage{hyperref} 
\makeatletter
\newcommand{\prlhead}[2][]{%
  \par\medskip\noindent
  \def\@currentlabelname{#2}%
  \phantomsection
  \if\relax\detokenize{#1}\relax\else\label{#1}\fi
  \textit{#2—}\ignorespaces
}
\newcommand{\prlsubhead}[2][]{%
  \par\smallskip\noindent
  \def\@currentlabelname{#2}%
  \phantomsection
  \if\relax\detokenize{#1}\relax\else\label{#1}\fi
  #2:\ \ignorespaces
}
\makeatother

\definecolor{awesome}{rgb}{0.2, 0.6, 0.99}

\newcommand{\cz}[1]{#1}

\newcommand{\Msun}{\ensuremath{M_\odot}}
\newcommand{\Mbh}{\ensuremath{M_{\rm BH}}}
\newcommand{\rsink}{\ensuremath{r_{\rm sink}}}

\newcommand{\rB}{\ensuremath{r_{\rm B}}}
\newcommand{\cs}{\ensuremath{c_s}}
\newcommand{\alphat}{\ensuremath{\alpha_{\rm target}}}
\newcommand{\alphar}{\ensuremath{\alpha_{\rm sink}}}

\newcommand{\rcore}{\ensuremath{r_c}}
\newcommand{\rhosol}{\ensuremath{\rho_{\rm sol}}}
\newcommand{\mboson}{\ensuremath{m}}

\newcommand{\kappaBoost}{\ensuremath{\kappa}}

\begin{document}

\title{Variability in Supermassive Black-Hole Accretion Rates in Fuzzy Dark Matter Cores due to Black-Hole Wandering}

\author{Eric Ludwig\,\orcidlink{0000-0003-0888-0789}}
\affiliation{Rensselaer Polytechnic Institute, Department of Physics, Applied Physics, \& Astronomy, 110 8th Street, Troy, NY 12180, USA}
\affiliation{Center for Computational Astrophysics, Flatiron Institute, 162 5th Avenue, New York, NY 10010, USA}

\author{Philip Mocz\,\orcidlink{0000-0001-6631-2566}}
\affiliation{Center for Computational Astrophysics, Flatiron Institute, 162 5th Avenue, New York, NY 10010, USA}

\author{Victor H. Robles\,\orcidlink{0000-0002-9497-9963}}
\affiliation{Rensselaer Polytechnic Institute, Department of Physics, Applied Physics, \& Astronomy, 110 8th Street, Troy, NY 12180, USA}

\begin{abstract}
Soliton cores in fuzzy dark matter (FDM) deepen nuclear potentials and have been proposed to strongly boost Bondi accretion, potentially aiding rapid black-hole growth at high redshift. We test this in \emph{live} Schr\"odinger--Poisson FDM cores coupled to isothermal gas, evolving a moving BH that grows via a strictly mass-conserving sink. We measure boosts relative to the initial mean-density Bondi rate. Low-mass seeds ($M_{\rm BH,init}\!\lesssim\!10^{6}\,M_\odot$) do not sustain large boosts: BH wandering and soliton sloshing drive bursty accretion with dense gas only intermittently present near the BH. Intermediate seeds ($M_{\rm BH,init}\!\sim\!10^{7}\,M_\odot$) produce the most durable enhancement, reaching $\mathcal{O}(10^{2})$ boosts for sound speed $c_s=60~{\rm km\,s^{-1}}$, \cz{while hotter gas approaches near-background Bondi rates}. High-mass seeds \cz{($M_{\rm BH,init}\!\sim\!10^{8}\,M_\odot$)} quickly exhaust the sink-scale reservoir and become supply-limited, suppressing long-lived growth despite the deepened potential. \cz{In general, central-potential deepening (e.g., by a soliton halo) does not guarantee long-lived fueling: sustained} boosts emerge only when \cz{the black hole remains dynamically confined} within the dense nuclear gas region. \cz{Our results suggest that SMBH formation channels relying on soliton-enhanced accretion alone are unlikely to provide sufficient early growth.}
\end{abstract}


\maketitle

\prlhead[sec:intro]{Introduction}
The \textit{JWST} era has rapidly expanded the census of accreting black holes at $z\gtrsim 10$ and luminous quasars with inferred dynamical masses $\gtrsim10^9\,\Msun$ already in place by $z\sim7$ \citep{BanadosJWST2018, YANGJWST2020, Larson2023, Kokorev2023, Bogdan2024, Kovacs2024JWST, Looser2024}. Proposed pathways for producing the initial black-hole seeds span Pop~III remnants \citep{Bromm2013, Sugimura2020}, heavy seeds via direct collapse \citep{Wise2019, Sassano2021, Mayer2024}, and intermediate-mass seeds from runaway stellar mergers \citep{Omukai2008, Devecchi2009}. The challenge is not merely seed formation, but sustaining the \emph{fueling history} needed to reach these masses on a short cosmic timeline: light-seed scenarios generally demand long duty cycles near (or above) the Eddington rate \citep{Smole2015, Inayoshi2016, Pezzulli2017, Mayer2019,Inayoshi2022}, while heavier seeds still require rapid growth that exceeds simple Bondi-type estimates \citep{Jeon2023, Jeon2025}. Regardless of the seeding channel, assembling a $\gtrsim10^9\,\Msun$ supermassive black-hole (SMBH) at high redshift requires efficiently delivering gas to sub-kiloparsec scales in environments that are turbulent, time-dependent, and strongly shaped by feedback \citep{DiMatteo2005,Hopkins2010,Dubois2015,Alcazar2017}. Many models invoke dense, long-lived nuclear reservoirs and steady inflow; whether such reservoirs form and persist in early galaxies remains uncertain \citep{Dubois2015, Alcazar2017}. A recurring issue is that simple spherical accretion scalings \citep{BHL3} can underpredict the supply needed unless the central gas density remains extremely high and the BH-gas relative velocities remain low for extended periods.

Ultra-light dark matter (ULDM) \citep{Guzm1999, Peebles2000, Matos2000, Bohmer2007, Sikivie2009, Hui2017, Hui2021} scenarios offer a qualitatively different dark-sector framework. In these models, the dark matter is a bosonic field with an extremely small mass, typically in the range $m \sim 10^{-22}$--$10^{-6}\,\mathrm{eV}$ and often realized as an axion-like particle. Such ultra-light fields arise naturally in high-energy theory, for example in constructions related to the quantum chromodynamics axion \citep{Wilczek1978} and in string theory compactifications \citep{Arvanitaki2010}, which can naturally yield a spectrum of light pseudo-scalar (axion-like) degrees of freedom spanning many orders of magnitude in mass. Within this broader class, the mass range $m_{\rm FDM} \sim 10^{-22}$--$10^{-20}\,\mathrm{eV}$ is commonly referred to as \emph{fuzzy} dark matter (FDM) \citep{Hu2000}, where the associated de Broglie wavelength reaches astrophysical (kiloparsec) scales and the wave nature of the field becomes dynamically important on halo scales. 
\textcolor{black}{Ultra-faint dwarf galaxies place important constraints on this parameter space, with some analyses favoring substantially heavier boson masses \citep{DalalKravtsov2022}. However, the interpretation of these constraints depends on the modeling of satellite stellar dynamics and in capturing nonlinear subhalo-host system evolution in self-consistent ULDM halos \citep{Chan2025}. Recent simulations suggest that Segue~1-like systems can arise at $m = 10^{-22}$~eV as near-disruption remnants, with their observed surface-density and velocity-dispersion profiles broadly matched in selected models once the nonlinear interplay between ULDM-induced heating, internal stellar relaxation, and stripping of outer stars is accounted for \citep{Yang2026}. Other dwarf-galaxy studies likewise have found agreement with FDM masses in the range considered here, depending on the modeling assumptions \citep{Chen2017, Wardana}.} On these scales, the same wave dynamics that suppress small-scale structure \citep{Marsh2016,Hui2017} also imply inherently time-dependent, non-particle-like halo behavior \citep{Mocz2017BecDM, Dutta2021, Dutta2023}, motivating treatments in which the central potential and local gas environment are not assumed to be static. For a recent overview of FDM phenomenology and Schr\"odinger--Poisson simulation methods, see \cite{Ferreira2021, Schive2025}.

In the fuzzy regime, Schr\"odinger--Poisson simulations show that halos develop dense central soliton cores embedded in an interference-dominated envelope \citep{Schive2014Profile}. These coherent, ground-state-like cores can deepen the central potential and raise nuclear gas densities relative to comparable cold-dark-matter halos. In idealized calculations, this has been argued to enhance Bondi-like accretion onto embedded BHs: \cite{Chiu2025} model a static soliton potential coupled to an effectively infinite, homogeneous gas reservoir and report large accretion ``boost factors'' relative to the standard Bondi rate at fixed ambient density and temperature. Here we test whether such enhancements survive in a more realistic, time-dependent setting with self-consistent BH growth, BH/soliton motion, and a finite gas reservoir—i.e., whether soliton-driven boosting is generically \emph{sustained} or instead fragile once dynamics and supply are treated consistently.

We address this question with controlled numerical experiments. We evolve a coupled FDM--gas--BH system using the time-dependent Schr\"odinger--Poisson (SP) equations for the FDM field and isothermal hydrodynamics for the gas, allowing both the BH and the soliton to move self-consistently in the combined potential. We grow the BH with a strictly mass-conserving, fixed-radius sink prescription—gas removed within $r_{\rm sink}$ is added to the BH mass each timestep—following the FLASH sink-particle methodology \citep{FLASH, FlashUG}. Since the sub-$r_{\rm sink}$ accretion flow is not resolved in our simulations, we use this standard sink-particle formalism as a robust subgrid closure that conserves mass by construction and cleanly isolates supply limitation within the resolved control volume. This enables a direct, time-resolved comparison between the analytic Bondi--Hoyle--Lyttleton ``target'' rate and the mass-conserving sink accretion rate.

Our aims are to (1) reinterpret soliton-induced boosted-Bondi expectations once BH-gas dynamics are treated consistently; (2) identify simple control parameters that determine when soliton-deepened potentials materially assist BH growth. We find that relative motion of BH-soliton wandering and finite gas supply jointly regulate both the instantaneous analytic accretion demand and the mass-conserving sink accretion.

\prlhead[theory]{Theory Background}
We evolve the self-gravitating Schr\"odinger--Poisson (SP) system for the FDM field $\psi$:
\begin{align}
i\hbar\,\partial_t \psi(\mathbf{x},t)
&= -\frac{\hbar^2}{2\mboson}\nabla^2 \psi
+ \mboson\,\Phi\,\psi,
\label{eq:sp-sch}
\end{align}
with gravity sourced by FDM, gas, and a BH in a periodic domain with mean-subtracted densities:
\begin{equation}
\label{eq:poisson}
\begin{gathered}
\nabla^2 \Phi
= 4\pi G\big[\rho_{\rm FDM}+\rho_{\rm gas}+\rho_{BH}
-\langle \rho_{\rm tot}\rangle\big],\\[2pt]
\rho_{\rm FDM} \equiv |\psi|^2,\\[2pt]
\rho_{BH}(\mathbf{x},t)\equiv
\big[\rho_{{BH},{\rm CIC}}(\mathbf{x},t)\star G_{r_{\rm sink}}(\mathbf{x})\big],\\[2pt]
\int \rho_{BH}\,d^3x = M_{BH}(t).
\end{gathered}
\end{equation}

Here $\rho_{{BH},{\rm CIC}}$ is the cloud-in-cell (tri-linear) deposition of the BH mass
$M_{BH}(t)$ onto the mesh at position $\mathbf{x}_{BH}(t)$ (nonzero only on the eight
neighboring cells, with weights summing to unity), and $G_{r_{\rm sink}}$ is a Gaussian
sink-scale kernel of width $r_{\rm sink}$ applied via Fourier filtering,
$\tilde\rho_{BH}(\mathbf{k})=\tilde\rho_{{BH},{\rm CIC}}(\mathbf{k})\exp[-k^2 r_{\rm sink}^2/2]$.
In our runs we take $r_{\rm sink}=2.5 \Delta x$, with $\Delta x$ being the cell size, and 
$M_{BH}(t)=M_{\rm seed}(t)+M_{\rm acc}(t)$, where the accreted mass, 
$M_{\rm acc}$, is determined by the gas mass removed from the sink each timestep. Over the $0.5\,{\rm Gyr}$ window used to measure accretion, this corresponds to $\simeq 5.5\times10^{4}$ timesteps, with a typical step $\Delta t_{\rm eff}\simeq 9.1\times10^{-6}\,{\rm Gyr}$.

The soliton core solutions are described by
\begin{equation}
\rhosol(r)=\rho_0\!\left[\,1+0.091\left(\frac{r}{\rcore}\right)^2 \right]^{-8},
\label{eq:soliton-fit}
\end{equation}
\citep{Schive2014Profile}, where $r_{c}$ is the core radius and $\rho_{0}$ is the central density given by
\begin{equation}
\rho_0 \simeq 3.1 \times 10^{15}
\left(\frac{2.5 \times 10^{-22}~\mathrm{eV}}{\mboson}\right)^{2}
\left(\frac{\mathrm{kpc}}{\rcore}\right)^{4}
\frac{M_\odot}{\mathrm{Mpc}^3}.
\end{equation}
Numerical simulations and analytic arguments relate the soliton to its host halo via approximate core--halo scalings, e.g.\ $r_c \propto m_{22}^{-1} M_h^{-1/3}$ and $M_{\rm sol}\propto m_{22}^{-1} M_h^{1/3}$ at fixed redshift, where $m_{22}\!\equiv\!\mboson/10^{-22}\,\mathrm{eV}$ \citep[e.g.][]{Schive2014corehalo,Chiu2025}. For halo masses $M_h \sim 10^{10}$--$10^{11}\,\Msun$ at $z \sim 7-9$ and $\mboson\sim 10^{-22}\,\mathrm{eV}$, this implies central solitons of $M_{\rm sol}\sim 10^7$--$10^9\,\Msun$ and kiloparsec-scale cores, so the soliton can carry a significant fraction of the inner mass budget and substantially deepen the central potential. 

In fully relaxed FDM halos, the soliton does not remain perfectly fixed at the halo center: interference with the surrounding turbulent envelope drives a stochastic center-of-mass ``random walk'' with an amplitude of order the core radius and characteristic timescales of order the core dynamical time \citep{Mocz2017BecDM}. Thus, even in the absence of any embedded BH, the central potential well is time-dependent on dynamical timescales, and any BH embedded in the core can experience both soliton motion and internal structural oscillations \citep{Mocz2017BecDM,Dutta2021}.

\prlsubhead[subsec:boost-defs]{Boost definitions}
For gas with ambient density $\rho_\infty$, sound speed $\cs$, and BH--gas relative velocity $v_{\rm rel}$, we adopt the classical Bondi--Hoyle--Lyttleton accretion rate \citep{BHL1,BHL2,BHL3}; for brevity, we refer to this expression as the ``Bondi'' rate throughout this work.
\begin{equation}
\dot M_{\rm target}
= 4\pi \lambda
\frac{(G\,\Mbh)^2 \rho_\infty}{(\cs^2+v_{\rm rel}^2)^{3/2}},
\label{eq:mdot-target}
\end{equation}
Here $G$ is Newton's gravitational constant, $\Mbh$ is the instantaneous BH mass, $\cs$ is the (isothermal) sound speed of the gas, and $v_{\rm rel}$ is the BH--gas relative speed. The quantity $\rho_\infty$ denotes the \emph{ambient} gas density entering the Bondi estimate; in our simulations we estimate $\rho_\infty$ from the resolved flow in a thin shell just outside the sink/control radius $\rsink$. The parameter $\lambda$ is a dimensionless factor of order unity that depends on the equation of state and the specific Bondi prescription. In what follows we refer to $\dot M_{\rm target}$ as the \emph{analytic target rate}: it is the instantaneous Bondi rate that the local flow \emph{demands}, based on the measured ambient density and relative velocity.

To factor out the global thermodynamics and the simulation box mean gas content, we compare all rates to a single, spatially homogeneous reference value evaluated at the initial box-mean gas density $\bar{\rho}_{{\rm gas},0}$ and the chosen sound speed. We denote this as the \emph{background} rate,
\begin{equation}
\dot M_{\rm bg}(\bar{\rho}_{{\rm gas},0},\cs;\Mbh)
\equiv 4\pi \lambda
\frac{(G\,\Mbh)^2 \bar{\rho}_{{\rm gas},0}}{\cs^3},
\label{eq:mdot-bg}
\end{equation}
which is just the Bondi rate for a BH of mass $\Mbh$ embedded in a uniform medium of density $\rho_{\rm gas}$ and sound speed $\cs$.

Our primary boost quantity is the \emph{analytic} boost,
\begin{equation}
\alphat \equiv
\frac{\dot M_{\rm target}}{\dot M_{\rm bg}(\bar{\rho}_{{\rm gas},0},\cs;\Mbh)} ,
\label{eq:alpha-target-def}
\end{equation}
which measures how strongly the local density and velocity field due to the soliton would enhance (or suppress) the accretion rate relative to the homogeneous background. 

The simulations also provide the \emph{sink} accretion rate $\dot M_{\rm sink}$, defined as the strictly mass-conserving rate obtained from the gas actually removed inside the sink/control volume each step. We first summarize how well the flow keeps up with the analytic demand by defining 
\begin{equation}
\kappaBoost \equiv \frac{\dot M_{\rm sink}}{\dot M_{\rm target}},
\label{eq:kappa-def}
\end{equation}
so that $\kappaBoost\!\approx\!1$ means that the flow supplies essentially the full analytic target rate, while $\kappaBoost\!<\!1$ indicates that the sink rate falls short (for example, because the gas inside the sink radius has been depleted and the flow has become explicitly supply-limited).

Using $\alphat$ and $\kappaBoost$, it is convenient to define a corresponding \emph{sink} boost,
\begin{equation}
\alphar \equiv \kappaBoost\,\alphat
= \frac{\dot M_{\rm sink}}{\dot M_{\rm bg}(\bar{\rho}_{{\rm gas},0},\cs;\Mbh)} ,
\label{eq:alpha-real-def}
\end{equation}
which is the actual enhancement of the numerically realized flow above the same background rate. By construction, $\alphar \leq \alphat$ whenever $\kappaBoost \leq 1$.

To diagnose when the accretion becomes supply-limited in our closed-box experiments, we track the total gas mass of the reservoir in the domain,
\begin{equation}
M_{\rm res}(t)\equiv \int_{V_{\rm box}} \rho_{\rm gas}(\mathbf{x},t)\,d^3x,
\label{eq:mgasbox-def}
\end{equation}
and its normalized form $M_{\rm res}(t)/M_{\rm res}(t_{\rm 0})$ (Fig.~\ref{fig:figure3}). In practice, sustained departures with $\kappaBoost<1$ occur only once the global reservoir has been strongly depleted; prior to that point, the sink rate typically remains demand-limited with $\kappaBoost\simeq 1$.

\prlhead[methods]{Methods}
Our simulations isolate the coupled FDM--BH--gas system in a periodic box while retaining the ingredients most relevant for boosted accretion. We perform them with a new JAX-based Schr\"odinger--Poisson + hydrodynamics code built on the
\textsc{Jaxion} framework \citep{Mocz_Jaxion_2025}. In this work we extend \textsc{Jaxion} by (i) adding a moving BH sink particle that contributes to the total gravitational potential, and (ii) evolving an isothermal gas component on the same mesh that is gravitationally coupled to the FDM and allowed to accrete conservatively onto the BH inside a fixed-radius control volume.

We evolve a coupled FDM--gas--BH system in a periodic box; the numerical and physical parameters are summarized in \hyperref[parameters]{Simulation Parameters}
. The gas is evolved isothermally having an adiabatic index $\gamma$=$1$ and sound speed $\cs\in\{60,70\}\,\mathrm{km\,s^{-1}}$ (corresponding to $T\simeq 2.6\times10^{5} \mathrm{K}$ and $3.6\times10^{5}\,\mathrm{K}$).
In our fiducial volume the initial conditions correspond to $M_{\rm FDM,box}\simeq 10^{10}\,M_\odot$ and a
central soliton of total mass $M_{\rm sol}\simeq 2\times10^{9}\,M_\odot$.
Prior to BH injection, we generate the FDM halo by merging solitonic cores and evolving until it relaxes to a quasi-steady soliton+interference
(granular) state \citep{Mocz2017BecDM}. After injection, each run is evolved for $T_{\rm post}\simeq\SI{0.5}{\gigayear}$ and all accretion diagnostics
are measured over this post-injection interval.

This setup generalizes earlier static potential or fixed-BH-mass models (e.g., \cite{Palomares2025}) by allowing:
(i) genuine BH growth,
(ii) BH and soliton motion (random walk, sloshing, migration),
(iii) explicit tracking of the finite local reservoir.
It also parallels velocity-focused work \citep{Lancaster2020DynFric, Boey2024DynFric} by tying the effective Bondi rate to $\cs$, $v_{\rm rel}$, and the evolving density field in a live core. These are closed-box experiments: we do not impose cosmological inflow, cooling-driven resupply, or external fueling. Thus, depletion reflects exhaustion of the resolved reservoir rather than feedback-driven evacuation.

The BH position is advanced using the local gravitational acceleration from the total SP+gas potential, allowing migration via dynamical friction and stochastic excursions driven by potential fluctuations. These modulate sampled density and the BH--gas relative velocity.

Ambient quantities for the analytic Bondi rate, Eq.~\eqref{eq:mdot-target}, are taken from the resolved flow in a thin shell just outside $\rsink$. When the BH is near the soliton/gas density peak this samples the highest-density gas accessible at the resolved scale; BH wandering shifts the shell off-peak, typically lowering $\rho_\infty$ and increasing $v_{\rm rel}$, and thus modulates $\dot M_{\rm target}$. Mass removal occurs only inside $\rsink$, is capped by the available mass each step, and the removed mass is added to the BH, defining the sink-measured rate $\dot M_{\rm sink}$. This allows the analytic demand and the strictly mass-conserving rate to be compared at every snapshot.

At each snapshot, we evaluate the background Bondi rate
$\dot M_{\rm bg}(\bar{\rho}_{{\rm gas},0},\cs;\Mbh)$ from the initial
box-mean gas density and chosen sound speed, compute $\dot M_{\rm target}$
from Eq.~\eqref{eq:mdot-target}, and measure $\dot M_{\rm sink}$ from gas
actually removed inside $\rsink$. Eqs.~\eqref{eq:alpha-target-def}--\eqref{eq:mgasbox-def}
then give $\alphat$, $\alphar$, $\kappaBoost$, and
$M_{\rm res}(t)/M_{\rm res}(t_0)$. We also record $M_{\rm BH}$,
$\rho_\infty$, $\rB$, BH--soliton separation, and central densities;
$M_{\rm res}(t)/M_{\rm res}(t_0)$ identifies when the system becomes
globally fuel-limited.

\begin{figure*}[t]
    \centering
    \includegraphics[width=0.9\textwidth]{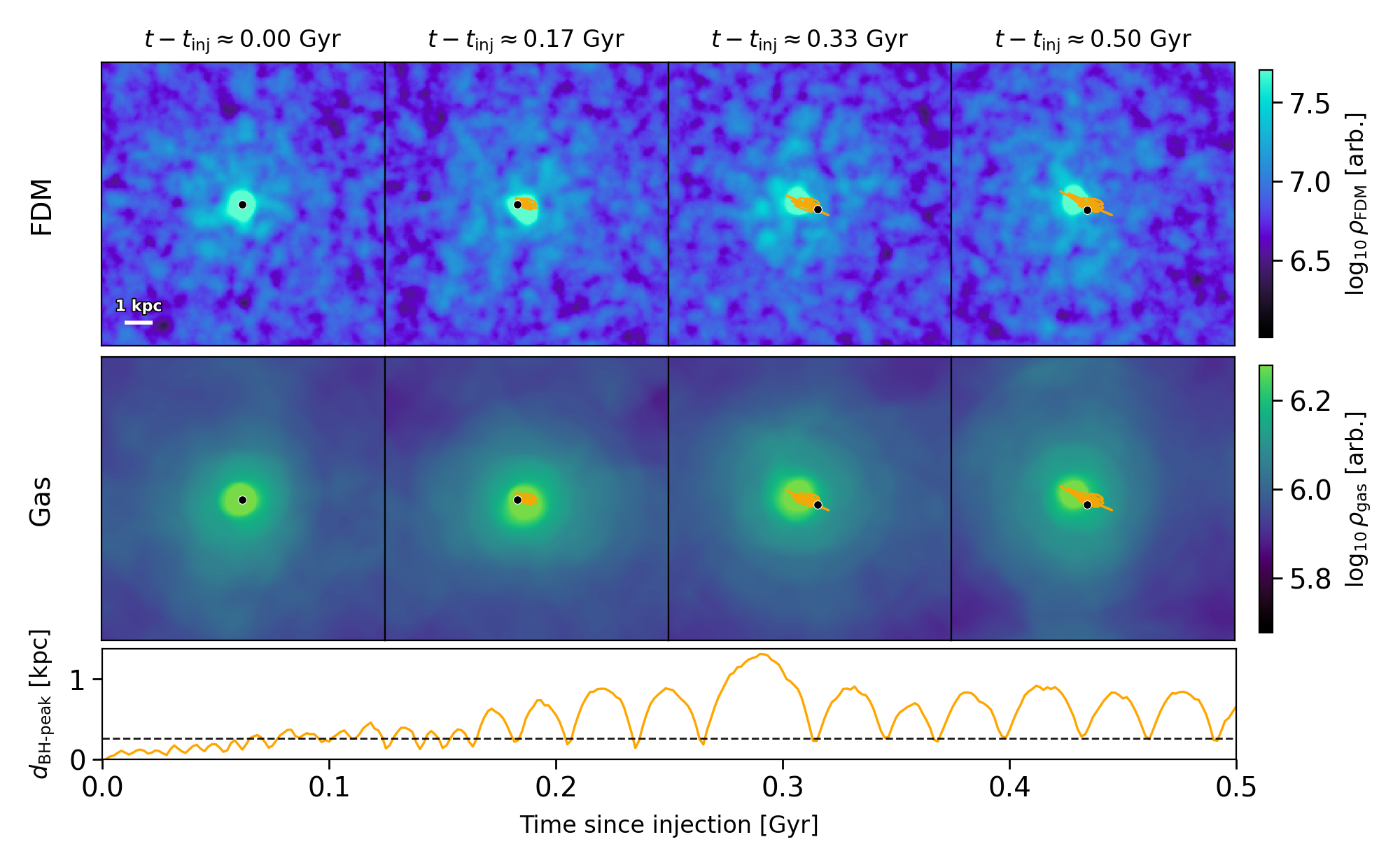}
    \caption{
    Shown: $c_s =  \{70\}\,{\rm km\,s^{-1}}$ and $M_{\rm BH,init} = 10^{6}\,M_\odot$ run; Black-hole injection into a live soliton core and subsequent BH--core wandering under the time-dependent FDM+gas potential.
    \emph{Top:} four midplane FDM density slices at times $t-t_{\rm inj}=\{0,\,0.17,\,0.33,\,0.50\}$~Gyr; the marker indicates the BH position and the line traces the BH trajectory up to each snapshot.
    \emph{Middle:} corresponding midplane gas density slices at the same times, shown with the same BH marker and trajectory overlay.
    \emph{Bottom:} distance between the BH and the instantaneous soliton density peak, $d_{\rm BH\text{--}peak}$, as a function of time since injection (dashed line indicates the median core radius $r_c$).
    The BH mass grows via a strictly mass-conserving sink operating inside the fixed control radius $r_{\rm sink}$, while the soliton and gas respond self-consistently to the evolving potential.
    }

    \label{fig:figure1}
\end{figure*}

\begin{figure*}[t]
    \centering
    \includegraphics[width=0.9\textwidth]{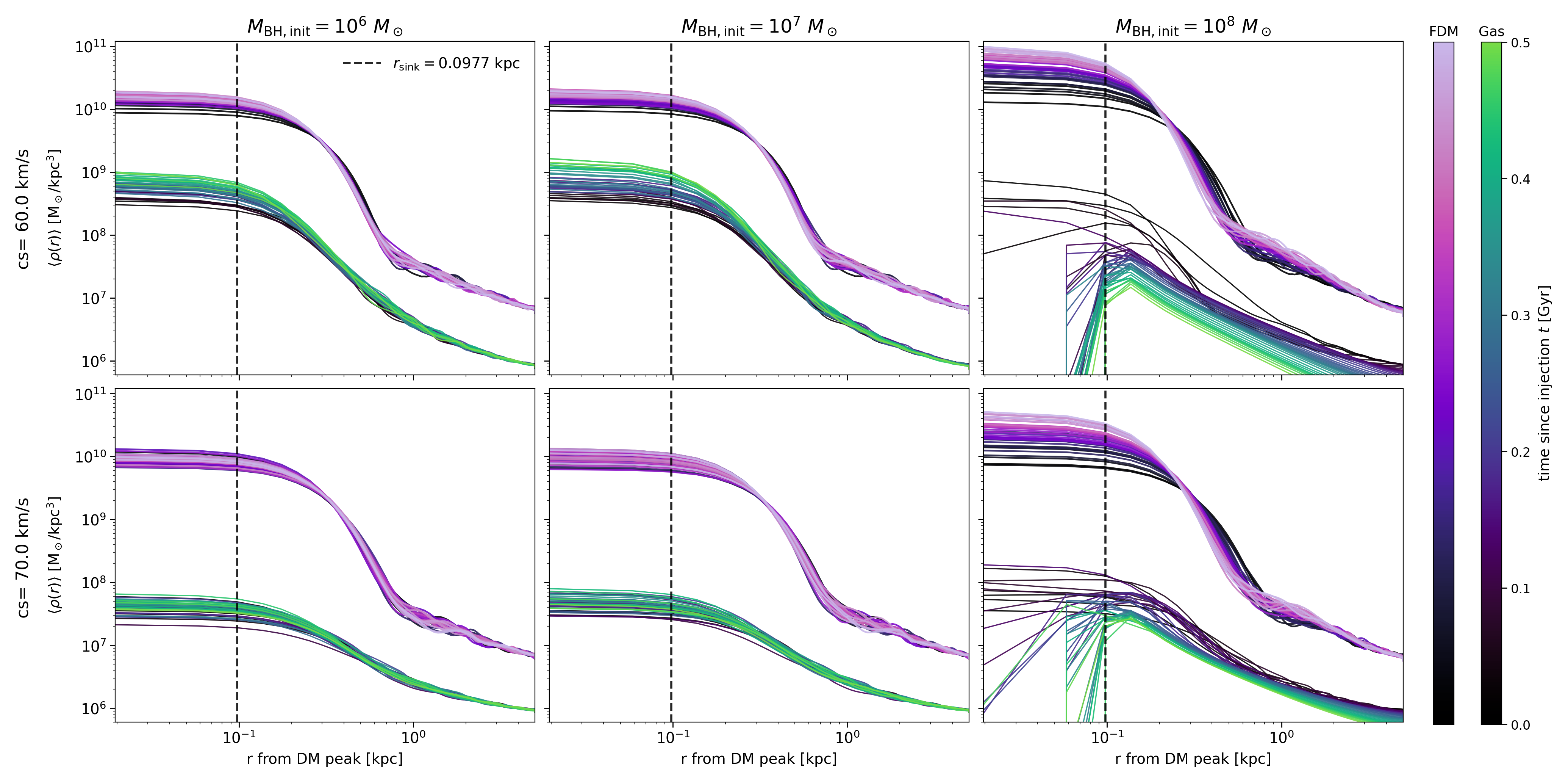}
    \caption{
    Radial density profiles of the FDM and gas measured about the instantaneous FDM density peak, illustrating thermodynamic contraction and BH-driven core evolution.
    Columns show the three BH seed masses ($M_{\rm BH,init}=10^6,\,10^7,\,10^8\,M_\odot$); rows show the two sound speeds ($c_s=60$ and $70~{\rm km\,s^{-1}}$).
    Within each panel, profiles are plotted at multiple times spanning the post-insertion evolution; the curve color encodes the simulation time (Gyr) as indicated by the accompanying colorbar.
    Lower-$c_s$ runs develop a more centrally concentrated gas+FDM configuration, while higher-mass seeds more strongly restructure the inner gas profile as accretion proceeds.
    }
    \label{fig:figure2}
\end{figure*}

\begin{figure*}[t]
    \centering
    \includegraphics[width=0.9\textwidth]{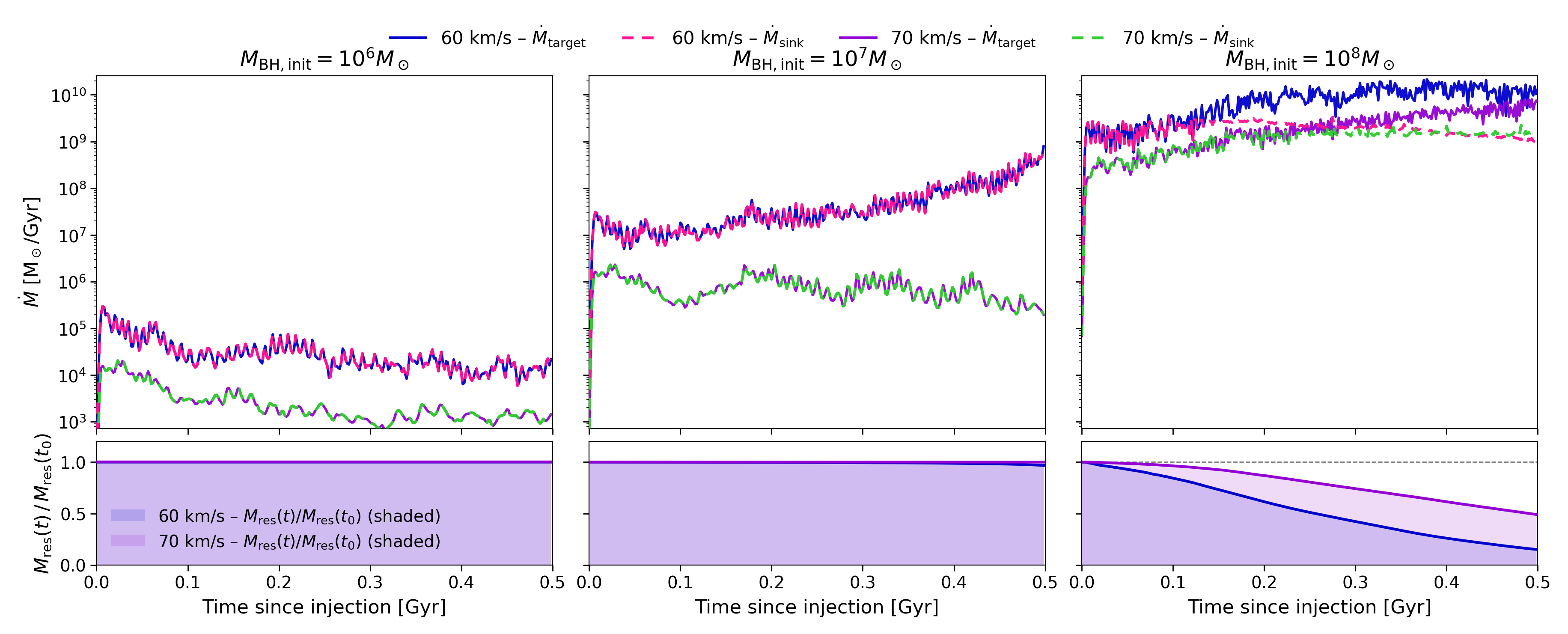}
    \caption{
    Black hole accretion histories and global gas reservoir for the six fiducial runs.
    Each column corresponds to a different initial BH mass
    ($M_{\rm BH,init}=10^{6},10^{7},10^{8}\,M_\odot$). 
    \emph{Top:} analytic accretion rate $\dot{M}_{\rm target}$ (solid) and mass-conserving sink rate $\dot{M}_{\rm sink}$ (dotted) versus time since injection, with both sound-speed runs ($c_s \approx 60$ and $70~\mathrm{km\,s^{-1}}$) overplotted in each panel.
    \emph{Bottom:} corresponding evolution of the total gas mass in the box, normalized to its value at injection, $M_{\rm res}(t)/M_{\rm res}(t_{\rm inj})$, again with both sound speeds overplotted; the shaded region indicates depletion relative to the initial reservoir.
    Low-mass seeds remain close to the target rate for most of the evolution, while the most massive BHs rapidly become globally fuel limited as the box reservoir is drained and $\dot{M}_{\rm sink} \ll \dot{M}_{\rm target}$.
    }

    \label{fig:figure3}
\end{figure*}

\begin{figure*}[t]
    \centering
    \includegraphics[width=0.9\textwidth]{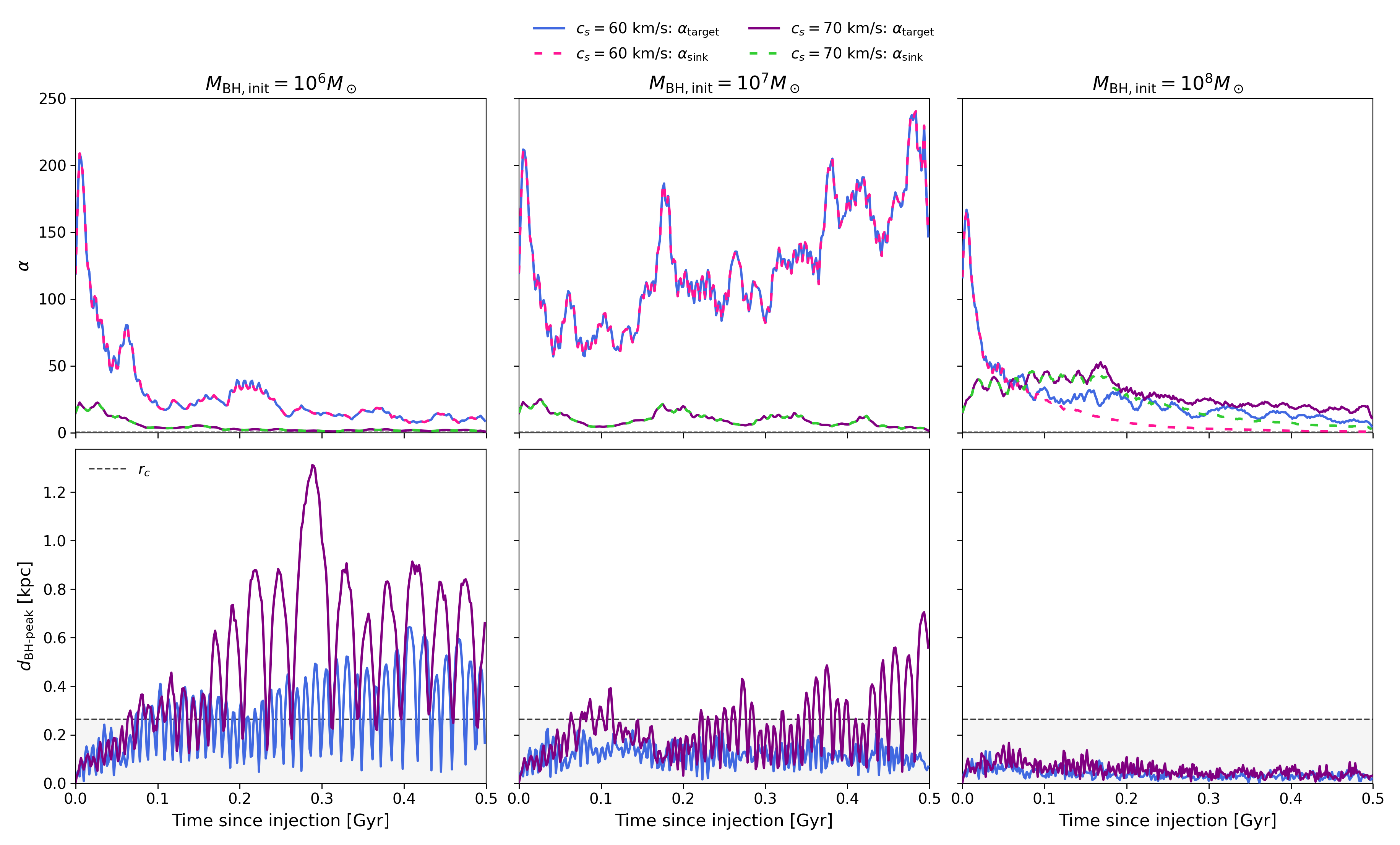}
    \caption{
    Boost evolution and BH--core separation for the same six fiducial runs and 2$\times$3 layout as Fig. \ref{fig:figure3}, with both sound-speed cases ($c_s \approx 60$ and $70~\mathrm{km\,s^{-1}}$) overplotted in each panel.
    In each \emph{top} panel, we show the analytic boost factor $\alpha_{\rm target}$ (solid; Eq.~\ref{eq:alpha-target-def}) and the sink boost $\alpha_{\rm sink}$ (dotted; Eq.~\ref{eq:alpha-real-def}) versus time since injection.
    In each \emph{bottom} panel, we show the BH distance from the instantaneous soliton density peak, $d_{\rm BH\text{--}peak}$; the horizontal dashed line marks the median soliton core radius $r_c$.
    Elevated boosts occur preferentially during intervals when the BH remains within (or repeatedly revisits) the soliton core, while sustained divergences between $\alpha_{\rm sink}$ and $\alpha_{\rm target}$ arise once local and/or global gas depletion sets in for the highest-mass runs.
    }
    
    \label{fig:figure4}
\end{figure*}

\begin{figure*}[t]
    \centering
    \includegraphics[width=0.9\textwidth]{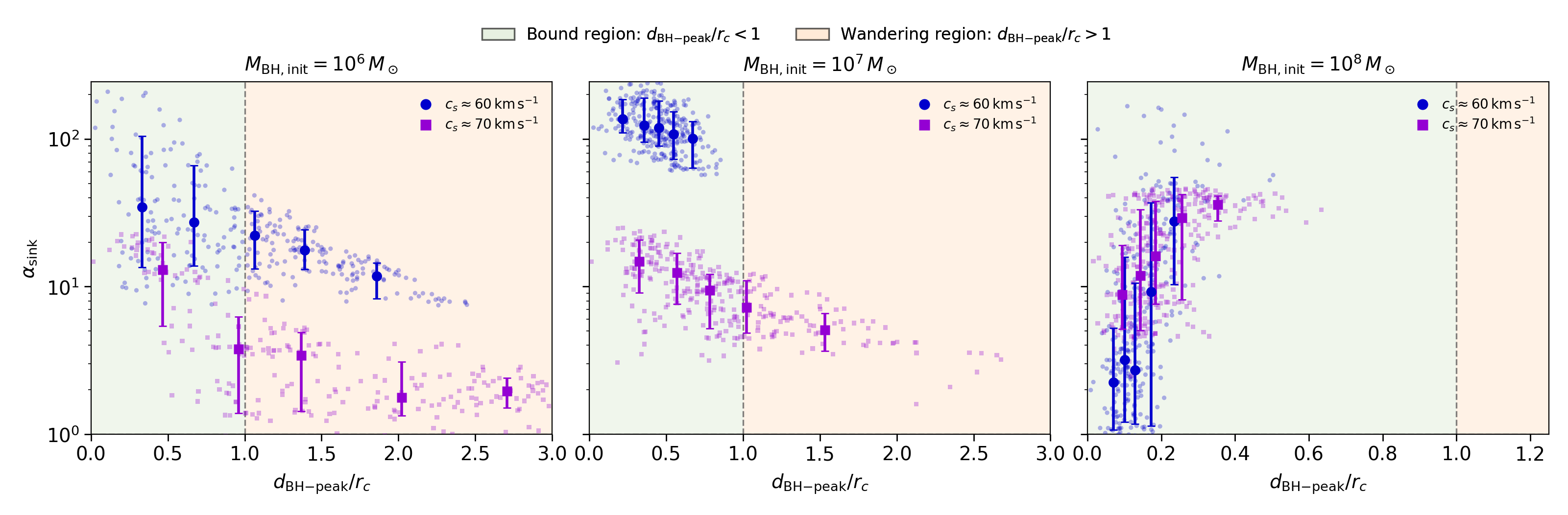}
    \caption{
    Sink boost as a function of BH offset from the soliton core.
    Each panel corresponds to a BH seed mass ($M_{\rm BH,init}=10^6,\,10^7,\,10^8\,M_\odot$).
    The x-axis shows the instantaneous BH--peak separation normalized by the soliton core radius, $d_{\rm BH\text{--}peak}/r_c$, and the y-axis shows the sink boost $\alpha_{\rm sink}$ (log scale).
    Small points show individual simulation snapshots; large markers show binned medians with error bars indicating the scatter within each bin (as defined in the text/legend).
    Shaded regions separate a “core-bound” regime ($d_{\rm BH\text{--}peak}/r_c<1$) from a “wandering” regime ($>1$; vertical dashed line).
    Across the suite, boosts are highest when the BH remains core-bound, and decline as the BH samples lower-density gas at larger offsets; the highest-mass case is additionally shaped by depletion-driven supply limitation.
    }

    \label{fig:figure5}
\end{figure*}

\prlhead[results]{Results}
For $\Mbh \lesssim 10^{7}\,\Msun$, accretion is demand-limited and boost variability is set primarily by dynamics. The analytic boost $\alphat$
fluctuates as the BH samples different $\rho_\infty$ and $v_{\rm rel}$ along its trajectory, and the sink boost $\alphar$ closely tracks
$\alphat$ with $\kappaBoost\simeq 1$ whenever gas remains available inside $\rsink$. Thus, variability in both boosts is dominated by BH motion and time-dependent soliton/gas structure rather than by supply limitation.

At fixed $\Mbh$, cooler gas runs ($\cs=60~\mathrm{km\,s^{-1}}$) attain systematically larger boosts than hotter gas, consistent with the Bondi scaling $\dot M \propto \cs^{-3}$.
In the $\Mbh=10^{6}\,\Msun$ cases, the BH can spend extended intervals outside the soliton core (often at $r \gtrsim r_c$), during which it samples lower densities and the boosts approach unity, $\alphat \simeq \alphar \simeq 1$; elevated boosts occur in intermittent episodes when the BH returns to the dense central region.
The $\Mbh=10^{7}\,\Msun$ seeds remain more tightly confined to the central region than the $10^{6}\,\Msun$ seeds, and therefore sustain higher typical boosts and longer high-$\alphar$ episodes in both sound-speed setups.

For $\Mbh \!\sim\! 10^{8}\,\Msun$, the target (Bondi) rate can become very large when the BH sits in dense gas with small $v_{\rm rel}$, producing large instantaneous $\alphat$ (and correspondingly large $\alphar$ while gas remains available inside the sink region).
A persistent separation between analytic and sink boosts occurs only after the gas reservoir inside $\rsink$ is drained:
as $M_{\rm res}$ declines, the sink rate can no longer satisfy the analytic demand, yielding $\alphar < \alphat$ and $\kappaBoost < 1$ (Fig.~\ref{fig:figure3}).
Before depletion, both $\alphat$ and $\alphar$ vary together in response to the same dynamical changes in density and $v_{\rm rel}$. Overall, boosted phases are intermittent and track BH core sampling, except in the $\Mbh\sim10^{8}\,\Msun$ runs where depletion drives sustained $\kappaBoost<1$.

\prlhead[discussion]{Discussion}
Our experiments connect static-potential soliton “boost” estimates with effectively infinite reservoirs \citep{Chiu2025} to
velocity-focused live-core dynamics \citep{Lancaster2020DynFric,Boey2024DynFric}. In a live FDM core, the enhancement is not set by the
potential depth alone: thermodynamics and dynamics jointly determine the instantaneous Bondi demand and whether the resolved flow can
continuously supply it.

At fixed $\Mbh$, lowering $\cs$ increases the analytic demand ($\dot M\propto c_s^{-3}$) and also alters the core state sampled by the BH.
The lower-$c_s$ runs develop more centrally concentrated FDM+gas cores (Fig.~\ref{fig:figure2}) and smoother boost histories
(Fig.~\ref{fig:figure4}), consistent with faster relaxation toward a quasi-steady central configuration. In particular for $\cs\sim70~\mathrm{km\,s^{-1}}$, the $\Mbh=10^{6}\,\Msun$ seed is more orbit-dominated and spends long intervals outside the dense core, so $\alphat$ and $\alphar$ often collapse toward unity.

Our parameter choice is also limited by numerical fidelity: we restrict to $\cs\in\{60,70\}\,\mathrm{km\,s^{-1}}$ because, in an isothermal
setup at fixed resolution, reducing $\cs$ shrinks the Jeans length $\lambda_J\propto \cs\,\rho^{-1/2}$ and would violate the Truelove
criterion \citep{Truelove1997}, driving artificial runaway central collapse. Across this range, the BH--gas relative speed is typically
$v_{\rm rel}\sim 40$--$150~\mathrm{km\,s^{-1}}$ for $M_{\rm BH,init}=10^{6}\,M_\odot$, $\sim 40$--$80~\mathrm{km\,s^{-1}}$ for
$10^{7}\,M_\odot$, and $\sim 25$--$60~\mathrm{km\,s^{-1}}$ for $10^{8}\,M_\odot$, with only weak dependence on $\cs$. Because
$\dot M_{\rm target}\propto \rho_\infty\,(c_s^2+v_{\rm rel}^2)^{-3/2}$, the $10^{6}\,\Msun$ runs often satisfy $v_{\rm rel}\gtrsim \cs$,
so further decreases in $\cs$ would yield only modest changes (diminishing returns once $\cs\ll v_{\rm rel}$) and would mainly amplify brief
near-core episodes rather than the long off-center intervals. For $10^{7}$--$10^{8}\,\Msun$ seeds, $v_{\rm rel}$ is typically smaller and
$\cs$ is comparable to or larger than $v_{\rm rel}$, so lowering $\cs$ can substantially increase the demand, both directly and indirectly
via stronger central contraction and enhanced BH confinement; in our closed-box setup, that larger demand also tends to hasten the onset of
explicit supply limitation once gas within $\rsink$ is depleted. Thus the $10^{6}\,M_\odot$ runs remain dynamics-dominated: lower $c_s$ would mostly boost only brief near-core phases (e.g., the initial peak),
not the long orbit-dominated intervals off-center.

At fixed thermodynamics, varying $\Mbh$ controls both core sampling and depletion: increasing $\Mbh$ strengthens dynamical friction and
central confinement, but it also raises demand and can accelerate draining of the local reservoir. This yields an intermediate regime at
$\Mbh=10^{7}\,\Msun$, which remains more tightly core-bound than $10^{6}\,\Msun$ but does not rapidly drain gas inside $\rsink$ like
$10^{8}\,\Msun$; the effect is strongest for $\cs=60~\mathrm{km\,s^{-1}}$, where core contraction enhances confinement and produces the
longest high-$\alpha$ duty cycles among the non--supply-limited runs (Fig.~\ref{fig:figure4}).

Finally, comparing sink accretion to analytic demand isolates when boosting is realized. While sufficient gas remains inside $\rsink$,
$\alphar$ closely tracks $\alphat$ with $\kappaBoost\approx 1$; sustained $\kappaBoost<1$ is tied to depletion of the sink control volume,
not an imposed ceiling on the prescription. In cosmological settings with sustained inflow from larger scales, the sink control volume could be continually refueled, so depletion-driven
$\kappaBoost<1$ may be less relevant and $\dot M_{\rm sink}$ would track the local demand until limited by galaxy-scale supply/feedback.
Moreover, our $10^{8}\,M_\odot$ seed in a fixed-mass core is intentionally idealized; establishing how such a massive BH would arise in this
environment requires separate modeling. Overall, while soliton-deepened potentials can support large instantaneous accretion demand in static-reservoir estimates,
in live FDM cores the sink boosts and their duty cycle are set by orbital sampling, soliton sloshing, and, in our closed box simulation, finite supply. 

\cz{Our simulations are isolated and omit several astrophysical ingredients that could shift the quantitative confinement and supply thresholds. Cosmological inflow or cooling-driven resupply could maintain the nuclear reservoir and keep \(\dot{M}_{\rm sink}\) closer to \(\dot{M}_{\rm target}\) for longer, while feedback, turbulence, angular momentum support, and galaxy-scale supply limits could instead shorten the lifetime of dense gas near the BH. Stars, nuclear star clusters, or additional compact objects could also enhance the effective dynamical friction acting on the BH. Previous high-resolution studies of massive black-hole seed dynamics have shown
that BHs embedded in surrounding stellar systems or nuclear star clusters can sink more efficiently toward the galactic center, whereas stripped BHs may stall at larger radii \citep{Tremmel2018WanderingBHs,Bellovary2021OffCenterDwarfs,Ma2021,Ogiya2020,Mukh2023,Zhou2025,Mukh2025}. In our context, including such stellar components could keep the BH more tightly confined to the soliton/gas center and thereby increase the high-boost duty cycle. However, modeling this self-consistently requires resolving the coupled gas, stellar, BH, and dark-matter dynamics in a cosmological environment, so both this setup and ours should ultimately be revisited with dedicated cosmological simulations. Such effects could allow lower-mass seeds to behave more like the confined cases studied here, but would not by itself guarantee SMBH assembly: the confined cases show that maintaining the boost is only one requirement, while the total mass gained by high redshift still depends on the gas thermodynamics and on how efficiently inflow maintains dense, low-relative-velocity gas around the BH.}

\prlhead[conclusions]{Conclusions}
In live, self-gravitating FDM soliton cores with conservative, fixed-radius sinks, accretion is regulated primarily by dynamics and becomes
explicitly supply-limited only after depletion in the most demanding cases. For $\Mbh\lesssim10^{7}\,\Msun$, boosts are intermittent because BH
motion controls how often the BH samples dense nuclear gas; in the hotter $\cs=70~\mathrm{km\,s^{-1}}$ case, the $10^{6}\,\Msun$ seed spends
extended intervals outside the core, so boosts frequently approach unity. In this low-mass regime, any accretion-powered emission could therefore
appear measurably off-nuclear, an observational consequence we leave to future work (e.g., \cite{Dokkum2023}). Sustained boosting is maximized
for intermediate seeds ($\Mbh\sim10^{7}\,\Msun$) in the cooler $\cs=60~\mathrm{km\,s^{-1}}$ runs, where confinement is strong without rapid local
depletion. This behavior is not simply a consequence of the BH and soliton masses being comparable: for $M_{\rm sol}\simeq2\times10^{9}\,\Msun$ we have
$M_{\rm BH}/M_{\rm sol}\in\{5\times10^{-4},\,5\times10^{-3},\,5\times10^{-2}\}$ for $\Mbh\in\{10^{6},10^{7},10^{8}\}\,\Msun$. When supply
limitation occurs, it is depletion-driven: only the $\Mbh\sim10^{8}\,\Msun$ runs sustain $\kappaBoost<1$ after gas inside $\rsink$ is exhausted.
Overall, deep soliton potentials can raise instantaneous demand, but large, long-lived boosts are not generic in live cores.

A key implication when combined with the soliton picture of \cite{Schive2014corehalo,Schive2014Profile} is that FDM can plausibly provide the \emph{potential} needed for large instantaneous Bondi demand in high-$z$ halos, but \cz{soliton-enhanced accretion alone is unlikely to provide sufficient early SMBH growth}: if the initial seed is too light to remain core-bound, wandering suppresses the duty cycle of high-density sampling and the boost collapses toward unity. In that case, the central question shifts to how the BH reaches the confinement-favorable mass scale (e.g., via heavy seeding or an early growth phase not captured by a closed-box setup), after which soliton deepening can help sustain high demand only if nuclear gas is continuously replenished.

\paragraph{\cz{Implications for cored dark-matter halos.}}
\cz{Large Bondi-like boosts require more than a deep central potential. The BH must remain co-located with the densest nuclear gas, and that gas must persist long enough to supply the implied demand.} Our live FDM experiments provide a concrete example: even with a deep central potential, boosts become intermittent when the BH spends extended intervals outside the core, \cz{as wandering lowers the surrounding gas density near the BH and raises \(v_{\rm rel}\)}, and become explicitly supply-limited only after the gas inside \(r_{\rm sink}\) is depleted.

\cz{The co-location requirement can arise in models where the central mass distribution is shallow or cored. In such systems, the restoring force and the phase-space structure that controls dynamical friction can differ substantially from a cuspy halo, allowing inspiral to slow or stall before the BH remains embedded in the nuclear gas. Constant-density cores are well known to exhibit weakened or stalled dynamical friction in collisionless systems \citep{Read2006,Goerdt2006,Petts2015,Petts2016}. FDM is a particularly strong realization of this issue because its soliton combines a flat inner density profile with intrinsic time dependence from wave interference, random walk, and granular fluctuations, which can further perturb BH orbits \citep{Schive2014Profile,Mocz2017BecDM,Dutta2021,BarOr2019RelaxationFDM,Hui2017}.}

\cz{SIDM therefore provides an important cored-halo comparison, although it is not a one-to-one analogue of FDM. SIDM halos can form extended low-density cores, and simulations have found more off-center SMBHs and delayed SMBH growth in such systems \citep{Dicint2017,Cruz2020SIDMSMBH}. At the same time, SIDM is not a one-to-one analogue of FDM: self-interactions can modify the core phase-space structure, reduce or eliminate core stalling in some regimes, and drive core contraction or gravothermal evolution \citep{TulinYu2018SIDM,DattathriVandenBosch2025,Jiang2025LRDsSIDM}. Thus the quantitative wandering amplitudes and boost duty cycles measured here should not be applied unchanged to SIDM, but the central requirement remains relevant: any boosted-accretion scenario in a cored halo must show both a mechanism that concentrates gas and a mechanism that keeps the BH co-located with that gas for a sufficiently long duty cycle.}

\paragraph{Software}: This work made use of the following software packages: \texttt{JAX} \citep{jax2018github}, \texttt{numpy} \citep{numpy}, \texttt{python} \citep{python}, \texttt{scipy} \citep{2020SciPy-NMeth}, \texttt{matplotlib} \citep{Hunter:2007}, \texttt{h5py} \citep{collette_python_hdf5_2014}.

Software citation information aggregated using \texttt{\href{https://www.tomwagg.com/software-citation-station/}{The Software Citation Station}} \citep{software-citation-station-paper,software-citation-station-zenodo}.

\prlhead[data]{Data Availability}
The code used in this work is available at \href{https://github.com/pmocz/quantum-jax}{github.com/pmocz/quantum-jax}. Simulation outputs and derived data products are available from the corresponding author upon reasonable request.

\begin{acknowledgments}
Eric Ludwig acknowledges support from the Center for Computational Astrophysics Pre-doctoral Program and the NASA New York Space Grant Fellowship. Simulations were performed on the Flatiron Institute Rusty computing cluster. The Flatiron Institute is supported by the Simons Foundation.
\end{acknowledgments}


\appendix

\prlhead[parameters]{Simulation Parameters}

\begin{table}[t]
\caption{\label{tab:params}Simulation parameters and run grid.
\textit{Note:} The suite comprises $6$ runs spanning the $(\cs,\,M_{\rm BH,init})$ grid.
Here $\Delta x$ denotes the grid cell width.}
\begin{ruledtabular}
\begin{tabular}{lll}
Parameter & Symbol & Value(s) \\
\hline
Boson mass & $m_\psi$ & $10^{-22}\,\mathrm{eV}$ \\
Grid resolution & $N$ & $256^{3}$ \\
Box size & $L$ & $10\,\mathrm{kpc}$ \\
Mean total density & $\bar{\rho}_{{\rm tot},0}$ & $10^{7}\,M_\odot\,\mathrm{kpc}^{-3}$ \\
FDM fraction & $f_{\rm FDM}$ & $0.9$ \\
Gas fraction & $f_{\rm gas}$ & $0.1$ \\
Sound speed & $\cs$ & $\{60,70\}\,\mathrm{km\,s^{-1}}$ \\
BH seed mass & $M_{\rm BH,init}$ & $\{10^{6},10^{7},10^{8}\}\,M_\odot$ \\
Sink radius & $r_{\rm sink}$ & $2.5\,\Delta x$ \\
Post-injection duration & $T_{\rm post}$ & $0.5\,\mathrm{Gyr}$ \\
Typical timestep & $\Delta t_{\rm eff}$ & $9.1\times10^{-6}\,\mathrm{Gyr}$ \\
Timesteps in $T_{\rm post}$ & $N_{\rm step}$ & $5.5\times10^{4}$ \\
\end{tabular}
\end{ruledtabular}
\end{table}


\bibliographystyle{apsrev4-2}
\bibliography{apssamp}

\end{document}